# A Reconfigurable Nanophotonics Platform for Sub-Millisecond, Deep Brain Neural Stimulation

Aseema Mohanty[1,2]*, Qian Li[3]*, Mohammad Amin Tadayon[1], Gaurang Bhatt[1], Euijae Shim[1], Xingchen Ji[1,2], Jaime Cardenas[1,4], Steven A. Miller[1], Adam Kepecs[3], and Michal Lipson[1]

[1] Department of Electrical Engineering, Columbia University, New York, New York 10027, USA
[2] School of Electrical and Computer Engineering, Cornell University, Ithaca, New York 14853, USA
[3] Cold Spring Harbor Laboratory, Cold Spring Harbor, New York 11724, USA
[4] Institute of Optics, University of Rochester, Rochester, New York 14627, USA
* These authors contributed equally to this work.

Nanophotonics, or chip-scale optical systems, provide the ability to rapidly and precisely reconfigure light beams on a compact platform. Infrared nanophotonic devices are widely used in data communications to overcome traditional bandwidth limitations of electrical interconnects[1–4]. Nanophotonic devices also hold promise for use in biological applications that require visible light, but this has remained technically elusive due to the challenges of reconfiguring and guiding light at these smaller dimensions[5–14]. In neuroscience, for example, there is a need for implantable optical devices to optogenetically stimulate neurons across deep brain regions with the speed and precision matching state-of-the-art recording probes[15–25]. Here we demonstrate the first platform for reconfigurable nanophotonic devices in the visible wavelength range and show its application *in vivo* in the brain. We demonstrate an implantable silicon-based probe endowed with the ability to rapidly (< 20 µs) switch and route multiple optical beams using a nanoscale switching network. Each switch consists of a silicon nitride waveguide structure that can be reconfigured by electrically tuning the phase of light and is designed for robustness to fabrication variation, enabling scalable, multi-functional devices. By implanting an 8-beam nanoprobe in mouse visual cortex, we demonstrate *in vivo* the ability to stimulate identified sets of neurons across layers to produce multi-neuron spike patterns and record them simultaneously with sub-millisecond temporal precision. This nanophotonic platform can be scaled up and integrated with high-density neural recording technologies, opening the door to implantable probe technologies that are able to simultaneously record and stimulate the activity of large neural populations at distant regions of the brain with sub-millisecond precision[1,15,16]. We expect this platform will enable researchers to test a variety of hypotheses about how spatio-temporal patterns of neural activity underlie behaviour and gain a deeper understanding into the neural code.

Nanophotonic devices are miniaturized and reconfigurable optical circuits with the potential to probe and control biological systems. These optical devices rely on micron-size tunable waveguide elements to control the phase of light and enable switching, steering, and shaping of optical beams using interferometric structures[2,5–8,26,27]. As a consequence, nanophotonic devices are fast and low power, in contrast to traditional table-top optical systems, which rely on moving parts that limit their speed due to their mass and size[28]. Nanophotonic devices can also be dramatically scaled up and integrated with electronics to create multifunctional optoelectronic microsystems since they are manufactured using standard nanofabrication techniques utilized by the electronic CMOS industry[1,13]. Nanophotonic modulators, switches, and reconfigurable filters have been demonstrated with speeds beyond a gigahertz and up to 1024 interconnected components, enabling minimal footprint for optical communication transceivers in the near-infrared region[29,6,30]. These optical transceivers have overcome fundamental bandwidth limitations of traditional electrical interconnects allowing for faster data transmission over long distances[1,4]. In addition, these miniaturized optical circuits have enabled new fields such as integrated quantum optics and optical phased arrays for remote sensing using LiDAR[9,31]. However, most reconfigurable nanophotonic devices have been developed for longer wavelengths in the near-infrared spectral range, limiting their applicability in biology, where visible spectral wavelengths (400-700 nm) are typically required (i.e. for neural control, fluorescent probes, spectroscopy and sensing, etc.)[32]. This is because of the high susceptibility of nanophotonic structures to fabrication imperfections in this short wavelength range, because the size of imperfections begin to approach the size of the optical mode in these high confinement structures[13,14,33].

In neuroscience, there is a need for an implantable optical stimulation technology for targeting neurons across deep regions of the brain with high spatial resolution and sub-millisecond timing precision shown to be critical to understanding neural circuits[21,22]. The compact size of today's implantable electrical recording probes enables observing neural activity from any region of the brain, including deep brain regions, during chronic behavioral experiments in freely moving animals[15]. High-density silicon probes based on CMOS electronic technology can measure the spiking activity of large neural populations at single-neuron resolution with sub-millisecond precision. Electrical recordings have revealed millisecond precision spiking in auditory, somatosensory, and motor systems, however, only perturbation studies, in which neural spikes are induced, can establish whether the information that is available at these time scales is actually used for network function and behavior[21–25,34–36]. Optogenetic techniques provide this ability to perturb specific neural circuit elements with light[37]. However, an optical technology is still needed for localized and ultrafast stimulation that can be integrated with electrical probes for deep brain accessibility. Micro-LEDs have been fabricated alongside integrated electrical probes, but they have limited spatial (>100 μm) resolution due to their incoherent Lambertian emission and limited temporal resolution (<30 Hz) due to heating and electrical interference effects[18–20]. Passive, implantable nanophotonic probes have been demonstrated that do not introduce these undesirable effects, however, they rely on external table-top optics for reconfiguration that limit the number of controllable optical stimulation sites and their chronic implantation[10–12,17,38–40]. Miniaturized lasers can also be integrated with nanophotonic waveguides, however, their assembly, size, and power dissipation limit the number that can be realized on a single probe[41]. The ability to reconfigure nanophotonic devices would enable light input from a single laser to be distributed to many highly localized emitters to create high resolution spatio-temporal optical patterns without introducing heating and electrical artifact effects.

To meet this demand, we demonstrate a scalable technology for enabling microsecond reconfigurable nanophotonic systems in the visible spectral range (at 473nm) that overcome the typically low fabrication tolerances of nanophotonic devices operating in this spectral range. The technology is based on waveguides (200 nm by 350 nm cross-section) defined in silicon nitride,

SiN, a material that is transparent in the visible range[42]. These high confinement waveguides ensure high dispersion and, hence, high sensitivity and fast response times to small changes in material index with low bending losses. These characteristics facilitate scalable, multi-functional devices that are reconfigurable and miniaturized. Because silicon nitride can be fabricated using standard deposition and etching processes that are CMOS compatible, it allows nanophotonic systems to be fabricated on silicon substrates that can be thinned to < 20 µm. In order to enable visible photonic systems that are reconfigurable, we design interferometric switches as the basic building blocks for routing light on a chip (see Fig. 1). These 1x2 (one input and two output) switches can be cascaded so that light from a single laser can be routed to $2^n$ output grating emitters. In order to induce robustness to fabrication variation, we design switches based on multimode interferometer (MMI) structures for splitting the light within the interferometer. These switches, due to their dimensional insensitivity are more robust than traditional directional coupler-based interferometers (see inset of Fig. 1 and Supplementary Figure 1)[43,44]. For reconfiguration, the switches are based on waveguides that are co-fabricated with highly localized microheaters that induce a small local temperature change, which in turn induces a refractive index change in the SiN waveguide ($\frac{dn}{dT}$ ~ 4×10$^{-5}$ RIU/K). An induced refractive index change (i.e. phase change) to one of the interferometer arms determines the spatial interference pattern at the switch output when the two arms are combined. Depending on the electrical power applied to this analog switch, light can exit either of the two output ports. In order to fabricate these structures, we leverage our recently demonstrated platform based on high confinement silicon nitride photonic technology with ultra-low loss in the near-infrared spectral range[42]. The fabrication process of the platform is described in detail in the Methods and Supplementary Figure 2.

To characterize the ultimate spatial resolution and speed of our platform, we measure the extinction ratio and switching time of a single fabricated switch. We show that the measured extinction ratio, or the ON/OFF contrast ratio, between the two outputs of the switch is high, about 50:1 or 17 dB (see Fig. 2a and Methods). This is comparable to the extinction ratio achieved in more traditional devices operating in the near infrared wavelengths[29]. This high extinction ratio enables full control over the path of light on our platform to allow independent control of output beams even for closely spaced emitters (< 1 µm apart). The resulting high spatial resolution of stimulation is important to selectively activate individual neurons. The switching power, or the power required to switch the light completely from port 1 to port 2, is approximately 30 mW. We also show fast switching time scales, on the order of 20 microseconds, well below the typical neural activity time scales that will enable advanced switching schemes in larger nanophotonic systems. We measure the time that it takes to switch the optical state of port 1 as the heater switch is turned off, and the light switches from output port 2 to output port 1 (see Fig. 2b and Methods). These switches can be used as a building block for larger nanophotonic systems with the same high spatial resolution and speed.

To illustrate the power of this platform, we built an implantable nanophotonic probe with 8 independently switchable beams with microsecond reconfiguration times. Using this platform we can control the flow of light dynamically using a network of switches, and we can direct light out of plane as narrow beams using grating emitters that are located at a distance from the switching network. By cascading multiple 1x2 Mach-Zehnder Interference (MZI) switches, we create an 8-beam probe using a 1x8 switching network in which the waveguides lead to 8 grating emitters. Fig. 1 shows the schematic of the implantable probe based on the high confinement silicon nitride waveguide structures designed for optogenetic excitation in the blue wavelength range centered at 473 nm (see Methods). Figure 2c shows a microscope image and scanning electron microscope (SEM) image of the fabricated and packaged device. Light is launched into the probe through an optical fiber at the top, coupled into a single waveguide, and then routed through the switching network. The output of the switching network is sent to grating emitters located at the bottom of the probe that send the light outward to excite the neurons. Using long routing waveguides, we place the active photonics that switches outside of the brain, several millimeters away from the target neural activation area in order to minimize interference or heating effects. Note that the gratings can be designed to emit beams at different angles and with different amplitude and phase distributions for controlling the degree of collimation and beam shape. Here we design the gratings to emit beams perpendicular to the probe, with minimal divergence. We chose the grating size to be 20 μm x 20 μm (see Methods) to approximately match the size of a single neuron.

We show the ability of this platform to form dynamically reconfigurable optical patterns of highly collimated beams. Figure 3 shows reconfigurable light patterns obtained by applying different power configurations to the switching network. We show the grating emitters produce low divergence beams by imaging the probe from the side as it excites a fluorescent dye (see Fig. 3b and Methods). Based on near-field imaging of the grating emitter's aperture, we measure the divergence angle to be 2.2° transverse to the waveguide propagation and 3.75° along the waveguide propagation. In Supplementary Movie 1, we show a video of the 8 spots being illuminated independently with low cross-talk. We use an optical power density of approximately of 15 mW/mm$^2$ incident upon a neuron 100 μm away, which exceeds the optical intensity threshold for optogenetic neural excitation, which ranges from 0.4 mW/mm$^2$ to 10 mW/mm$^2$ in previous studies[18].

The nanophotonic probe enabled us to activate individual neurons with sub-millisecond precision *in vivo*. Figure 4 shows examples of electrically recorded signals from optically activated neurons across multiple brain regions. Neural activity was recorded with an array of tungsten electrodes aligned to the grating emitters of a nanophotonic probe inserted into the visual cortex and the hippocampus of anaesthetized mice (see Fig. 4a, Supplementary Figure 4 and Methods). We rendered inhibitory neurons light-activatable by injecting an adeno-associated virus delivering a channelrhodopsin-2 variant (AAV-EF1a-DIO-ChETA-eYFP) in transgenic knock-in mice (Gad2-IRES-Cre). We chose this mouse line because Gad2 inhibitory neurons locally inhibit other neurons, minimizing secondary light-activation of neurons not expressing the light-sensitive opsin. We used ChETA, a channelrhodopsin-2 variant, because it enables ultrafast optogenetic control[45]. Figure 4a shows that light-induced and spontaneously occurring spikes exhibited similar spike waveforms (spike waveform similarity, $r > 0.95$, see Methods), as expected from single neuron signals. Alignment to the 1 ms light pulse revealed low latency and low jitter spikes (Fig. 4b). Increasing light intensity above the threshold did not yield additional spikes due to secondary activation by other neurons. Based on the low latency and jitter of spikes, and because we targeted inhibitory neurons these responses are from directly light-activated neurons. We estimate approximately 1-2 inhibitory neurons can be activated within a 100 µm beam path (the approximately relevant radius for electrical recordings), based on the estimated optical volume and neuron density. We found that an individual light beam activated at most one neuron and occasionally we co-recorded putative pyramidal neurons with short-latency troughs in their cross-correlograms, indicative of monosynaptic inhibition from the recorded Gad2 interneuron (see Supplementary Figure 5). These results confirm the ability of the probes to activate individual neurons as observed by recordings from these activated neurons.

Using the 8-beam nanophotonic probe, we generated precisely timed multi-neuron spike patterns *in vivo*. Figure 4 shows the generation of different spatio-temporal neural patterns including sequential bursts, random pulses, and high-frequency patterns (up to 200 Hz). Figure 4c shows a representative trial of the recorded neural spike patterns that follow the optical command patterns produced by the beams (blue ticks). These patterns could be evoked with high reliability across neurons and multiple trials (Fig. 4d). We quantified reliability by computing the probability of light induced spike responses (peri-stimulus histogram), revealing >90% fidelity across all these patterns. Supplementary Figure 6 shows additional measurements of two neurons switched at multiple frequencies. The spike latency and jitter of light induced spikes across repeated patterns were consistently low for stimulation frequencies up to 200 Hz (Fig. 4e; n=13 neurons from 4 mice). The high-speed and reliable switching in these experiments demonstrate that our nanophotonic probe can be used *in vivo* to drive precise and high fidelity population spike patterns.

The demonstrated visible nanophotonic switch in the platform can be used as a fundamental unit for larger, multi-functional nanophotonic systems. These devices can be scaled up and improved using several approaches. The power dissipation can be reduced to the sub-milliwatt regime per switch using optimized microheater architectures[26,46,47]. To reduce the footprint of emitters along the inserted probe, one can multiplex many optical channels into a single waveguide using different degrees of freedom of light (i.e. wavelength or transverse spatial mode) to control multiple emitters[4]. Finally, the broadband nature of the photonic devices allows for multi-colour experiments involving optical stimulation and inhibition and targeting multiple neuron types simultaneously[32]. The nanophotonic waveguides are also compatible with multi-photon optical approaches, enabling further reach for deep-brain stimulation[21,48].

Our demonstration of a reconfigurable visible nanophotonic platform is the first step towards implantable optical probes with rapid control over arbitrary populations of single neurons for brain stimulation for neuroscience. While here we designed devices that control the phase and amplitude of light to produce fixed beams, the same platform can be extended to create arbitrary patterns of optical stimulation[5,27]. For example, optical phased arrays based on nanophotonic switches and grating emitters in the near-infrared wavelength range have been used to create steerable and focused beams[5–7]. Furthermore, the natural integration between silicon photonics and electronics will enable integrated probes with optical stimulation capacity that matches the scale and resolution of available electrical recording probes[1,12,15]. The development of such multifunctional large-scale devices will enable the control over arbitrary, genetically or functionally defined neural populations to study the contribution of precisely timed multi-neuron patterns to neural computation and behaviour.

## METHODS
### Nanophotonic probe design
Here we outline the details of the nanophotonic probe design from input fibre to output grating emitters following Fig. 1 from top to bottom. For fibre-to-chip coupling, we use a horn taper at the input to match the optical mode of a cleaved single mode fibre (460B). Next, the switching network for the 1x8 nanophotonic probe consists of seven thermally-tuned Mach-Zehnder Interferometer (MZI) switches (see Supplementary Figure 1 for more details on the switch design). We use EigenMode Expansion (Fimmwave) to numerically simulate the optical properties of the MZI switch and optimize the design for fabrication, wavelength, and polarization insensitivity. Each MZI is composed of a 300 µm long platinum microheater for phase control and two multimode interferometer (MMI) beamsplitters for interference. The waveguides throughout the switching network are based on waveguides with a cross-section of 200 nm height and 350 nm width. We chose this cross-section to minimize losses due to sidewall roughness while remaining near the second order mode cut-off at 473 nm. The routing waveguides from the switching network are adiabatically tapered to 700 nm for the mm-long routing waveguides and then to 20 µm at the grating emitters. We use fully etched grating emitters that are designed to produce beams that are near 90°. We use a grating pitch of 260 nm with a 50:50 duty cycle. The grating emitters span about 1 mm and are spaced 80 µm horizontally and 125 µm vertically.

### Nanophotonic probe fabrication and packaging
To fabricate the implantable nanophotonic probe, we deposit 200 nm of low-pressure chemical vapour (lpcvd) silicon nitride on a silicon wafer with 5 µm of lpcvd silicon dioxide. Next, we pattern the waveguides using electron beam lithography and a fluorine-based etch. We cover the waveguides with 660 nm of lpcvd high temperature oxide. Next, to pattern the switch above the waveguide, we use a metal lift-off process to pattern 100 nm of platinum with a titanium adhesion layer. We dice the chip at an angle, so that the inserted tip is less than 100 µm. Finally, the silicon substrate below the inserted tip is thinned to 250 µm using a partially masked Bosch etch process. The final tip dimensions are approximately 250 µm x 100 µm. The probe chip was wire-bonded to a PCB for controlling the switch, and a fibre was aligned to the chip and attached using UV-curable optical adhesives. For in vivo measurements, an array of tungsten electrodes (1 MOhm, World Precision Instruments) are aligned near the grating emitters and attached to the probe using permanent adhesive. We align the tungsten electrode tip around 20 µm away from the grating emitters. These steps are outlined in detail in Supplementary Figure 2.

### Nanophotonic probe performance characterization
To deliver light to the probe, we use a cleaved single mode fibre (460B) to input light from a blue diode laser centred at 473 nm (Model SSL-473-0300-10TM-D-LED, Shanghai Sanctity Laser Technology Co.). For electrical input to the switch, we use a multichannel digital-to-analog converter module (NI-DAQ) with an amplified output. For the extinction ratio measurement, we use a Newport detector (818-SL), and for the measurement of the time-transient, we use a biased silicon detector with 1 ns rise time (DET10A). We take images and video of the switching patterns using a Thorlabs CMOS camera. To image the beams from the side to characterize the beam profile, we insert the device in small glass container filled with fluorescent dye (Alexa-Fluor 488, 32 µM solution) and image using a camera (JVC TK-S241U).

### Animals for in-vivo experiments

Adult (6-8 months old) male or female Gad2-IRES-Cre knock-in mice (Jackson Labs, #010802) were used under the protocol approved by Cold Spring Harbour Laboratory Institutional Animal Care and Use Committee in accordance with National Institutes of Health regulations. Mice were maintained with a reverse 12 hr light/dark cycle with ad libitum food and water.

**Virus injection**

Gad2-IRES-Cre knock-in mice were anaesthetized with Ketamine and Xylazine mixture (100/10 mg/kg, IP) and placed into a stereotactic frame (SR-6M-HT, Narishige, Tokyo, Japan). A sagittal incision along the midline was made to expose the cranium, and a craniotomy was made over the target area, the visual cortex (V1) (AP: -2.6 mm, ML: 2.4 mm). 1 µl of Cre-dependent AAV9-EF1a-DIO-ChETA-eYFP vector (a ChR2 variant from UNC vector core, high frequency stimulation drivable), was loaded into a glass micropipette (tip diameter ~20 µm) attached to a syringe and was injected into V1 and below in hippocampus at four depths (DV: 0.2, 0.4, 0.8, 1.3 mm from brain surface, 0.25 µl per site, viral titer: $8 \times 10^{12}$ vg/ml). After each injection the micropipette was left in place for 5 minutes before it was retracted to a higher position. The scalp incision was sutured. The mouse was maintained under a 37°C heater and under observation until recovery from the anaesthesia, before returning to standard cages. Mice were maintained for 4 weeks before performing electrophysiological experiments. The delivery of large volume of AAV and the long waiting time for viral expression increased the efficacy and reduced the variability of ChETA expression.

*In vivo* **experiments**

For acute *in vivo* experiments, mice were anesthetized and placed in the same stereotactic frame as described for virus injection. The craniotomy for virus injection was exposed again and enlarged to a 2 x 2 mm$^2$. The dura was carefully removed and the 8-beam probe was inserted slowly using a motorized arm (Thorlabs MTS50-Z8) to target viral infected areas crossing layer 2-6 of the visual cortex (V1), with the deepest activation site reaching ~1300 µm from brain surface. For electrophysiological recordings, extracellular spike signals of 3 tungsten stereotrodes were simultaneously pre amplified (20X), filtered (band-pass, 600-6 KHz) and sampled at 30 kHz using the integrated Digital Lynx recording system (Neuralynx, Inc). For patterned light stimulation, we delivered 1 ms laser pulses at multiple frequencies (40 Hz, 80 Hz, 200 Hz) and patterns (regular bursting or random). A pulse generator (Sanworks LLC Pulse Pal v2)[49] was programmed by Matlab through USB serial interface, which sent 7-analog voltage outputs in different combinations to control the optical switch network and 1 synchronized TTL output to trigger the laser pulse. The timing for switch voltage and laser pulse were acquired by splitting the TTL output channel to the DI/O port of the recording system (see Supplementary Figure 3 for the experimental setup). We screened for neural activity based on an on-line peri-stimulus time histogram (HistogramDisplay V1.3.0, Neuralynx) by aligning spike events to the onset of each light pulse. Light induced spikes were detected electrophysiologically in vivo since illumination with a brief flash of blue light reliably elicited a short latency, temporal jittered spike. Single neuron identification was performed off-line (see below). In initial experiments we observed light-induced electrical artefacts, but these were abolished by eliminating the light leakage from the external fibre coupling. Supplementary Figure 7 shows light induced artefacts are very tightly time-locked to the light pulse onset, whereas real light induced spikes have more temporal jitter.

**Electrophysiological data analysis**

All data analysis was carried out using built-in and custom-built software in Matlab (Mathworks). Spikes were manually sorted into clusters off-line based on peak amplitude and waveform energy using the MClust software as described from previous literature[50]. We confirmed that optically tagged neurons were directly light-activated based on the stimulus-associated spike latency test

(SALT) in combination with the spike waveform similarity $(r)^{51}$. Here we define spike waveform similarity, $r$, as the correlation coefficient between the averaged waveform of light-induced and spontaneous spikes. Next, we generated spike raster and peri-stimulus time histogram (PSTH) with light stimulations delivered with multiple frequencies and patterns. To address the issue of spike fidelity, light-induced spike probability was defined as the proportion of successfully light induced spikes to light pulses; To quantify the light activation effect, light-induced spike latency was defined as the time from light pulse onset to the time of the spike that occurred after the onset of a light pulse; light-induced spike jitter was defined as the standard deviation, across repeated trials, of the timing of the spike that occurred after the onset of a light pulse. Direct light-activated neurons showed similar spontaneous and light-induced waveforms (high spike similarity, $r$ value), high light-induced spike probability, short light-induced spike latency, and low light-induced spike jitter.

**Histology**
Following acute electrophysiology recordings, animals were anesthetized and transcardially perfused with 0.9% saline followed by 4% paraformaldehyde in Phosphate Buffered Saline (PBS). Brains were extracted, immersed in 4% paraformaldehyde for 24 h, and then prepared for sectioning by washing with PBS for three times, each lasting 5 min. The brain tissue was embedded with 2% agarose gel, mounted on the vibratome stage (Leica VT 1000S) and sliced into 90-micron thick coronal sections including the V1 and hippocampus. The tungsten stereotrodes were coated with DiI labelling solution (VybrantTM) before surgery so that the insertion track could be verified histologically (see Supplementary Figure 4). The histological images were acquired with Zeiss LSM780 confocal laser scanning microscope.

**REFERENCES**

1. Atabaki, A. H. *et al.* Integrating photonics with silicon nanoelectronics for the next generation of systems on a chip. *Nature* **556,** 349–354 (2018).

2. Xu, Q., Schmidt, B., Pradhan, S. & Lipson, M. Micrometre-scale silicon electro-optic modulator. *Nature* **435,** 325–327 (2005).

3. Reed, G. T., Mashanovich, G., Gardes, F. Y. & Thomson, D. J. Silicon optical modulators. *Nature Photonics* **4,** 518–526 (2010).

4. Dai, D., Wang, J., Chen, S., Wang, S. & He, S. Monolithically integrated 64-channel silicon hybrid demultiplexer enabling simultaneous wavelength- and mode-division-multiplexing: Monolithically integrated 64-channel silicon hybrid demultiplexer. *Laser & Photonics Reviews* **9,** 339–344 (2015).

5. Sun, J., Timurdogan, E., Yaacobi, A., Hosseini, E. S. & Watts, M. R. Large-scale nanophotonic phased array. *Nature* **493,** 195–199 (2013).

6.  Hutchison, D. N. *et al.* High-resolution aliasing-free optical beam steering. *Optica* **3,** 887 (2016).

7.  Phare, C. T., Shin, M. C., Miller, S. A., Stern, B. & Lipson, M. Silicon Optical Phased Array with High-Efficiency Beam Formation over 180 Degree Field of View. *arXiv:1802.04624 [physics]* (2018).

8.  Shen, Y. *et al.* Deep learning with coherent nanophotonic circuits. *Nature Photonics* **11,** 441–446 (2017).

9.  Harris, N. C. *et al.* Quantum transport simulations in a programmable nanophotonic processor. *Nature Photonics* **11,** 447–452 (2017).

10. Segev, E. *et al.* Patterned photostimulation via visible-wavelength photonic probes for deep brain optogenetics. *Neurophotonics* **4,** 011002 (2016).

11. Shim, E., Chen, Y., Masmanidis, S. & Li, M. Multisite silicon neural probes with integrated silicon nitride waveguides and gratings for optogenetic applications. *Scientific Reports* **6,** 22693 (2016).

12. Li, B., Lee, K., Masmanidis, S. C. & Li, M. A Nanofabricated Optoelectronic Probe for Manipulating and Recording Neural Dynamics. *Journal of Neural Engineering* (2018). doi:10.1088/1741-2552/aabc94

13. Selvaraja, S. K., Bogaerts, W., Dumon, P., Van Thourhout, D. & Baets, R. Subnanometer Linewidth Uniformity in Silicon Nanophotonic Waveguide Devices Using CMOS Fabrication Technology. *IEEE Journal of Selected Topics in Quantum Electronics* **16,** 316–324 (2010).

14. Choy, J. T. *et al.* Integrated TiO_2 resonators for visible photonics. *Optics Letters* **37,** 539 (2012).

15. Jun, J. J. *et al.* Fully integrated silicon probes for high-density recording of neural activity. *Nature* **551,** 232–236 (2017).

16. Du, J., Blanche, T. J., Harrison, R. R., Lester, H. A. & Masmanidis, S. C. Multiplexed, High Density Electrophysiology with Nanofabricated Neural Probes. *PLoS ONE* **6,** e26204 (2011).

17. Pisanello, F. *et al.* Dynamic illumination of spatially restricted or large brain volumes via a single tapered optical fiber. *Nature Neuroscience* **20,** 1180–1188 (2017).

18. Wu, F. *et al.* Monolithically Integrated µLEDs on Silicon Neural Probes for High-Resolution Optogenetic Studies in Behaving Animals. *Neuron* **88,** 1136–1148 (2015).

19. Kim, T. -i. *et al.* Injectable, Cellular-Scale Optoelectronics with Applications for Wireless Optogenetics. *Science* **340,** 211–216 (2013).

20. Scharf, R. *et al.* Depth-specific optogenetic control in vivo with a scalable, high-density µLED neural probe. *Scientific Reports* **6,** 28381 (2016).

21. Shemesh, O. A. *et al.* Temporally precise single-cell-resolution optogenetics. *Nature Neuroscience* **20,** 1796–1806 (2017).

22. Peron, S. & Svoboda, K. From cudgel to scalpel: toward precise neural control with optogenetics. *Nature Methods* **8,** 30–34 (2011).

23. Yang, Y., DeWeese, M. R., Otazu, G. H. & Zador, A. M. Millisecond-scale differences in neural activity in auditory cortex can drive decisions. *Nature Neuroscience* **11,** 1262–1263 (2008).

24. deCharms, R. C. & Merzenich, M. M. Primary cortical representation of sounds by the coordination of action-potential timing. *Nature* **381,** 610–613 (1996).

25. Hahnloser, R. H. R., Kozhevnikov, A. A. & Fee, M. S. An ultra-sparse code underliesthe generation of neural sequences in a songbird. *Nature* **419,** 65–70 (2002).

26. Watts, M. R. *et al.* Adiabatic thermo-optic Mach–Zehnder switch. *Optics Letters* **38,** 733 (2013).

27. Cai, X. *et al.* Integrated Compact Optical Vortex Beam Emitters. *Science* **338,** 363–366 (2012).

28. Ji, N., Freeman, J. & Smith, S. L. Technologies for imaging neural activity in large volumes. *Nature Neuroscience* **19,** 1154–1164 (2016).

29. Tanizawa, K. *et al.* Ultra-compact 32 × 32 strictly-non-blocking Si-wire optical switch with fan-out LGA interposer. *Optics Express* **23,** 17599 (2015).

30. Phare, C. T., Daniel Lee, Y.-H., Cardenas, J. & Lipson, M. Graphene electro-optic modulator with 30 GHz bandwidth. *Nature Photonics* **9,** 511–514 (2015).

31. Poulton, C. V. *et al.* Coherent solid-state LIDAR with silicon photonic optical phased arrays. *Optics Letters* **42,** 4091 (2017).

32. Zhang, F. *et al.* The Microbial Opsin Family of Optogenetic Tools. *Cell* **147,** 1446–1457 (2011).

33. Zortman, W. A., Trotter, D. C. & Watts, M. R. Silicon photonics manufacturing. *Optics Express* **18,** 23598 (2010).

34. Wagner, H., Brill, S., Kempter, R. & Carr, C. E. Microsecond Precision of Phase Delay in the Auditory System of the Barn Owl. *Journal of Neurophysiology* **94,** 1655–1658 (2005).

35. Garcia-Lazaro, J. A., Belliveau, L. A. C. & Lesica, N. A. Independent Population Coding of Speech with Sub-Millisecond Precision. *Journal of Neuroscience* **33,** 19362–19372 (2013).

36. Bale, M. R., Campagner, D., Erskine, A. & Petersen, R. S. Microsecond-Scale Timing Precision in Rodent Trigeminal Primary Afferents. *The Journal of Neuroscience* **35,** 5935–5940 (2015).

37. Boyden, E. S., Zhang, F., Bamberg, E., Nagel, G. & Deisseroth, K. Millisecond-timescale, genetically targeted optical control of neural activity. *Nat Neurosci* **8,** 1263–1268 (2005).

38. Hoffman, L. *et al.* High-Density optrode-electrode neural probe using SixNy photonics for in vivo optogenetics. (2015).

39. Zorzos, A. N., Scholvin, J., Boyden, E. S. & Fonstad, C. G. Three-dimensional multiwaveguide probe array for light delivery to distributed brain circuits. *Optics Letters* **37,** 4841 (2012).

40. Zorzos, A. N., Boyden, E. S. & Fonstad, C. G. Multiwaveguide implantable probe for light delivery to sets of distributed brain targets. *Optics Letters* **35,** 4133 (2010).

41. Schwaerzle, M., Paul, O. & Ruther, P. Compact silicon-based optrode with integrated laser diode chips, SU-8 waveguides and platinum electrodes for optogenetic applications. *Journal of Micromechanics and Microengineering* **27,** 065004 (2017).

42. Ji, X. *et al.* Ultra-low-loss on-chip resonators with sub-milliwatt parametric oscillation threshold. *Optica* **4,** 619 (2017).

43. Thomson, D. J., Hu, Y., Reed, G. T. & Fedeli, J.-M. Low Loss MMI Couplers for High Performance MZI Modulators. *IEEE Photonics Technology Letters* **22,** 1485–1487 (2010).

44. Mikkelsen, J. C., Sacher, W. D. & Poon, J. K. S. Dimensional variation tolerant silicon-on-insulator directional couplers. *Optics Express* **22,** 3145 (2014).

45. Gunaydin, L. A. *et al.* Ultrafast optogenetic control. *Nature Neuroscience* **13,** 387–392 (2010).

46. Densmore, A. *et al.* Compact and low power thermo-optic switch using folded silicon waveguides. *Optics Express* **17,** 10457 (2009).

47. Fang, Q. *et al.* Ultralow Power Silicon Photonics Thermo-Optic Switch With Suspended Phase Arms. *IEEE Photonics Technology Letters* **23,** 525–527 (2011).

48. Horton, N. G. *et al.* In vivo three-photon microscopy of subcortical structures within an intact mouse brain. *Nat Photon* **7,** 205–209 (2013).

49. Sanders, J. I. & Kepecs, A. A low-cost programmable pulse generator for physiology and behavior. *Frontiers in Neuroengineering* **7,** (2014).

50. Pi, H.-J. *et al.* Cortical interneurons that specialize in disinhibitory control. *Nature* **503,** 521–524 (2013).

51. Kvitsiani, D. *et al.* Distinct behavioural and network correlates of two interneuron types in prefrontal cortex. *Nature* **498,** 363–366 (2013).

**ACKNOWLEDGEMENTS**
This work was supported by the National Science Foundation Brain EAGER (#1611090). This work was performed in part at the Cornell NanoScale Facility, a member of the National Nanotechnology Coordinated Infrastructure (NNCI), which is supported by the National Science Foundation (Grant ECCS-1542081). Back-end fabrication processing was done in part at the Advanced Science Research Center (ASRC) NanoFabrication Facility at the Graduate Center of the City University of New York. A.M. was funded by a National Science Foundation Graduate Research Fellowship (Grant No. DGE-1144153). X.J. acknowledges the China Scholarship Council for financial support.

**AUTHOR CONTRIBUTIONS**
A.M. designed and tested the performance of the nanophotonic probe. Q.L. conducted the animal surgery, histological analysis, and electrophysiology data analysis. A.M. and Q.L. developed and conducted the in-vivo electrophysiology experiment with the assistance of M.A.T. A.M. fabricated the nanophotonic probe with the assistance of X.J. and J. C. E.S. and G.B. assisted with back-end fabrication processing and electrical packaging, respectively.  A.M. and M.A.T. developed the fibre packaging for in-vivo experiments. S.A.M. developed the software interface for optical characterization. A.M., Q.L., M.A.T., A.K., and M.L. designed the experiment and discussed the results. A.M., Q.L., and M.A.T. designed and built the experimental set-up. A.K. and M.L. supervised the project. A.M., Q.L., A.K., and M.L. prepared the manuscript. M.A.T., G.B., E.S., X.J., J.C., and S.A.M. edited the manuscript.

# FIGURES

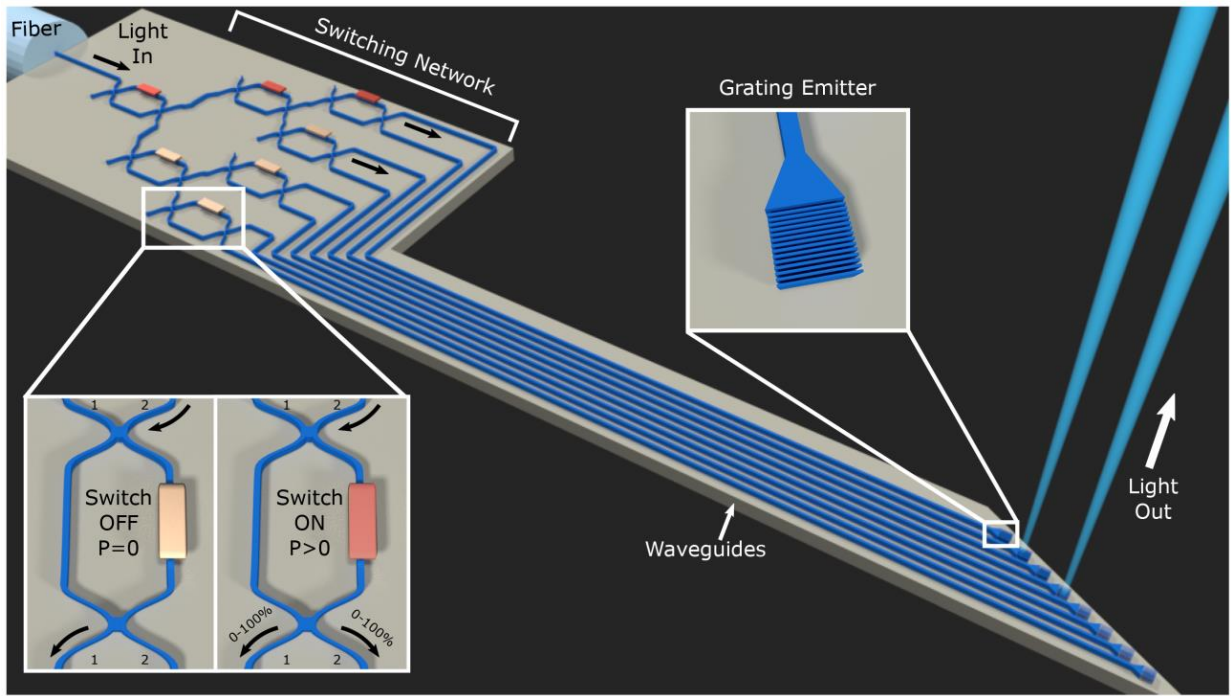

*Figure 1 Schematic of an implantable probe based on the reconfigurable nanophotonic platform operating in the visible spectral range for optogenetic neuromodulation.* The probe forms independently switchable beams within microsecond time scale. Light is input through a fiber at the top of the device into a single waveguide which is sent to the switching network. The output of the switching network is sent to grating emitters located at the bottom of the probe that send the light outward to excite the neurons. To minimize electrical interference and heating effects, we use long routing waveguides to place the active photonics that switches the light outside of the brain, several millimeters away from the target neural activation area. Bottom Left Inset: Shows a single 1x2 MMI-based switch that routes light between two ports. depending on the electrical power applied to the switch, light can exit any of the two output ports. If light enters the switch through input port 1 (2), if no power is applied (P = 0, left) light exits through output port 2 (1). As the power on the switch is increased (P > 0, right), light is continuously tuned between output port 1(2) and output port 2(1) from 0 to 100%, until all the light is completely in port 2(1). Top Right Inset: Schematic of grating emitter designed to direct light out-of-plane.

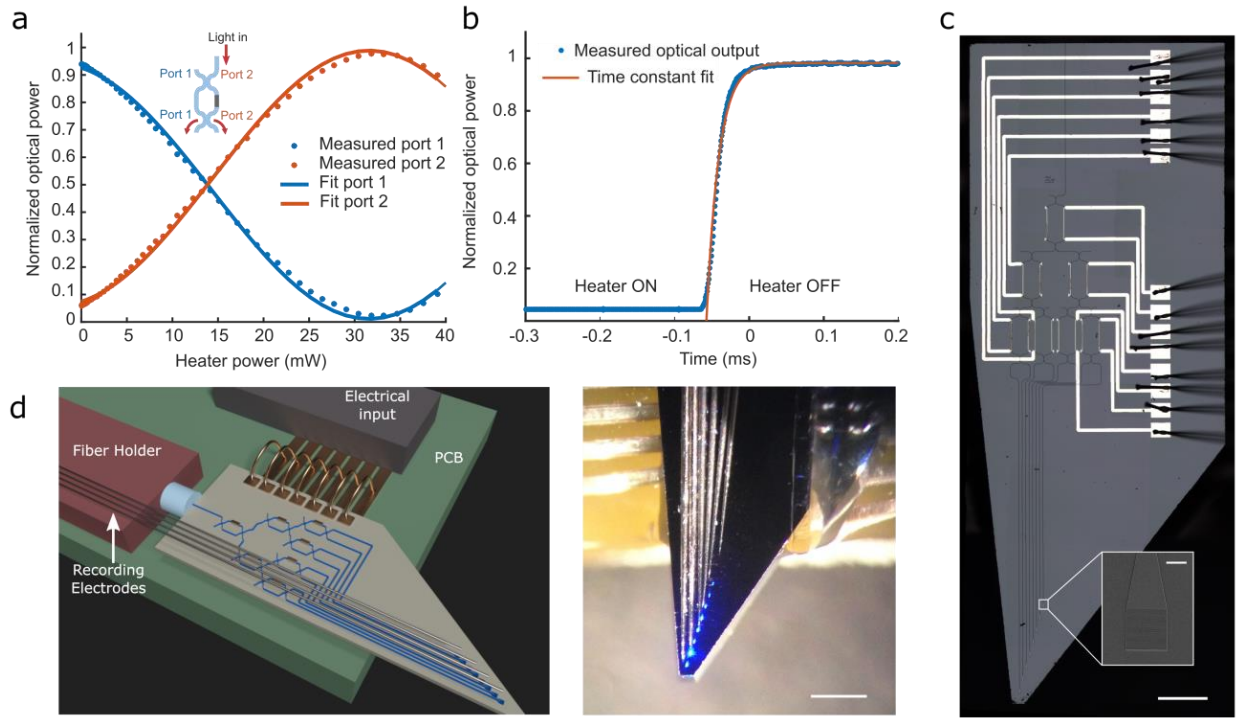

*Figure 2 Nanophotonic switch operating in the visible spectral range with high extinction ratio and short switching time.* a) *Normalized optical output power measured through output port 1 and output port 2 of the MZI switch, when 473 nm light is input through input port 2. For a single switch, the measured switch power for full conversion from output port 1 to output port 2 is 30 mW and the ON:OFF contrast ratio is 50:1 (or 17 dB extinction). b) Response time of output port 1 switching on when the heater is turned off. The switching time (τ) is 20 μs, which we calculate by fitting an exponential to the measured optical output. c) Microscope image of the fabricated chip (scale bar is 500 μm). Inset: scanning electron microscope image of the grating emitter (scale bar is 10 μm). d) Left: The schematic shows the fiber and electrical packaging of the neural probe. Right: Microscope image of the packaged device with electrodes aligned near the emitters (around 20 μm) and optical output (scale bar is 500 μm).*

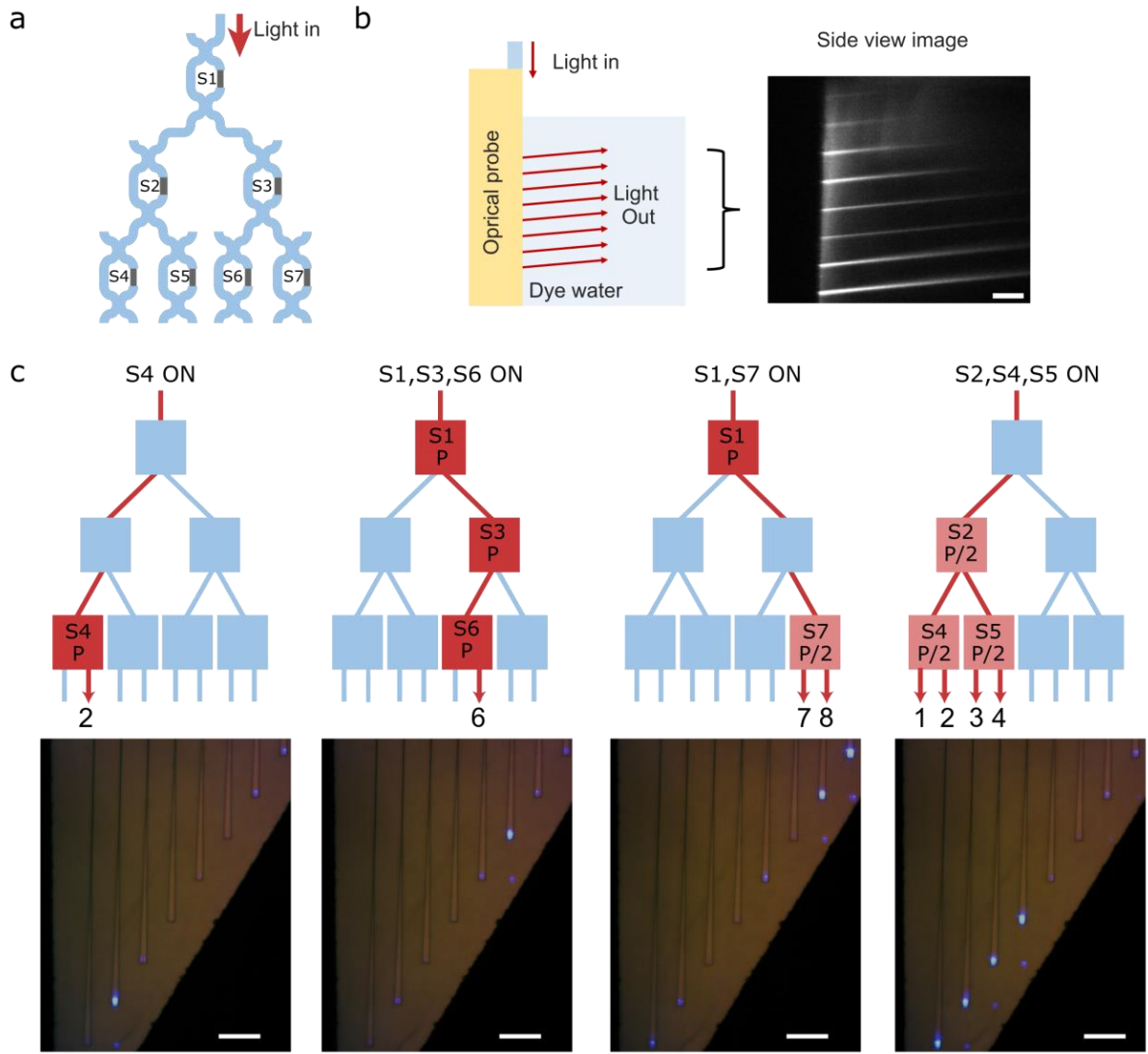

*Figure 3 Spatial optical patterns generated by the probe to demonstrate dynamic reconfiguration of highly collimated beams. a) Schematic of the optical switch network. b) Left: Schematic of setup to image the spatial distribution of light output from the nanophotonic probe in fluorescent dye from the side. Right: Side view of the grating emitters with 473 nm blue light transmitting through Alexa-Fluor 488 dye (scale bar is 125 μm). c) Examples of different switch configurations that produce different optical patterns (scale bar is 125 μm). Light enters the switch network through the top and exits through the arrows indicated below. Switches indicated by a blue box are OFF, and switches indicated by a red box are ON. P indicates full power and P/2 indicates half the power.*

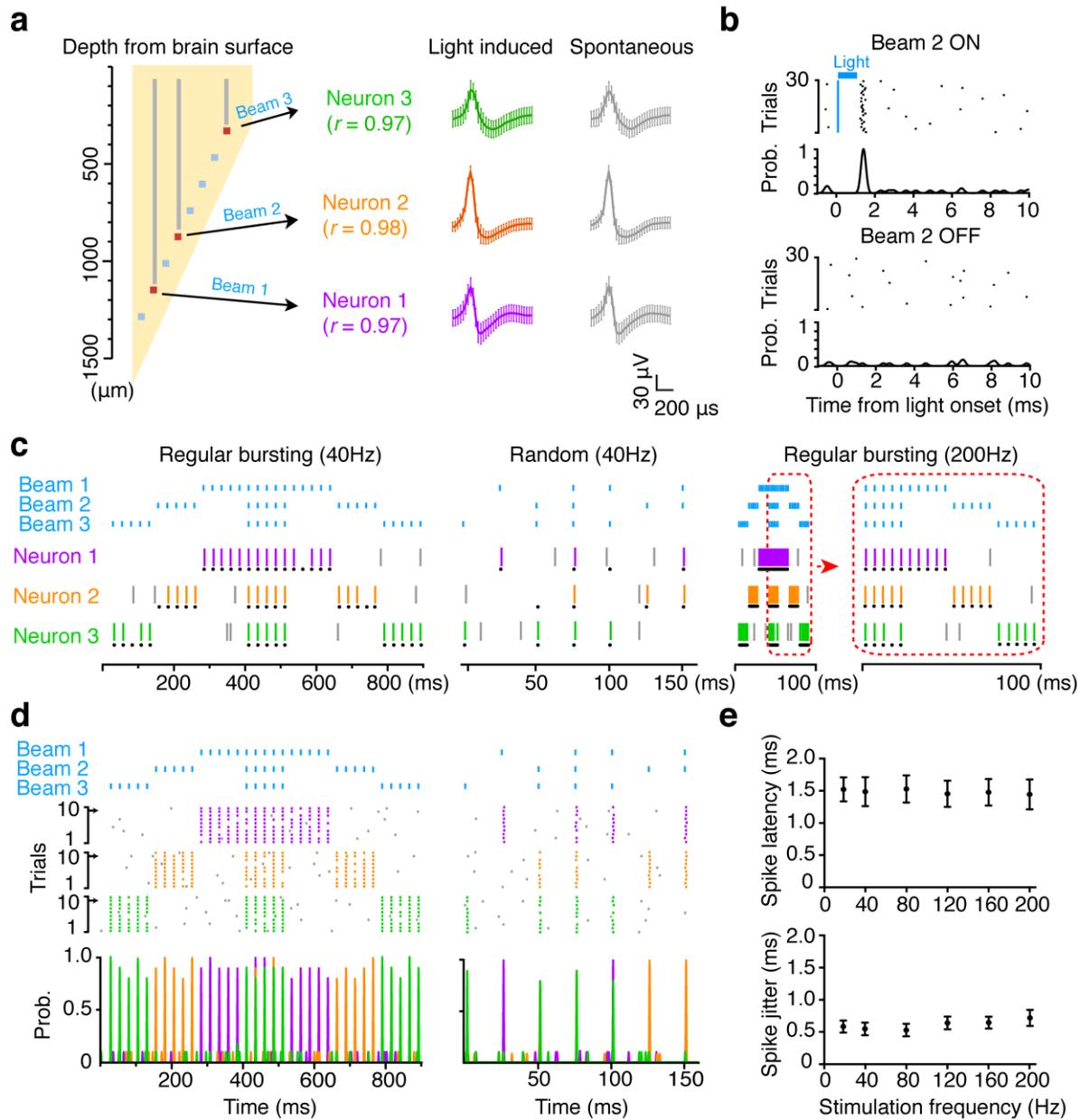

*Figure 4 In vivo demonstration of probe driving multiple individual neurons with precisely timed spike patterns.* a) Schematic of 8-beam probe for fast, independent control of three ChETA-expressing Gad2 interneurons' activities across layer 2-6 visual cortex (V1) and hippocampus in anesthetized mice. Three pairs of electrical recording stereotrodes (3 grey lines) are positioned next to beam 1-3 (3 red squares) respectively. From each recording site, single ChETA-expressing Gad2 interneuron (Neuron 1-3) is confirmed by evaluating the spike waveform similarity (r) of light induced spikes (averaged spike waveform in colored bold line, superimposed with s.d. of individual spikes in colored error bars) and spontaneous spikes (averaged spike waveform in grey bold line, superimposed with s.d. of individual spikes in grey error bars). b) Representative spike raster and peri-stimulus time histogram of Neuron 2 when the light beam 2 is ON and OFF. Blue line indicates 1 ms light pulse and pulse onset is set to time '0'. c) Spike raster of Neuron 1-3 driven by series of light pulses delivered in 3 patterns: 1) regular bursting sequence at 40 Hz, 2) random sequence with highest frequency at 40 Hz, and 3) regular bursting sequence at 200 Hz. Blue line indicates light pulse. For each neuron, colored line indicates light induced spike, grey line indicates spontaneous spikes, and black dot indicates the timestamp when light induced spike supposes to occur. d) Representative 10 trials of spike raster and peri-stimulus time histogram of Neuron 1-3 in response to repeated regular bursting (left) and random(right) light stimulation at 40 Hz. Bin size: 100 µs. Black arrows point to the specific trial shown in panel c. e) Summary of spike latency (top) and spike jitter (bottom) referring to light onset throughout multiple stimulation frequencies (mean ± s.d., n=13, sorted neurons from 4 mice).

# SUPPLEMENTARY FIGURES

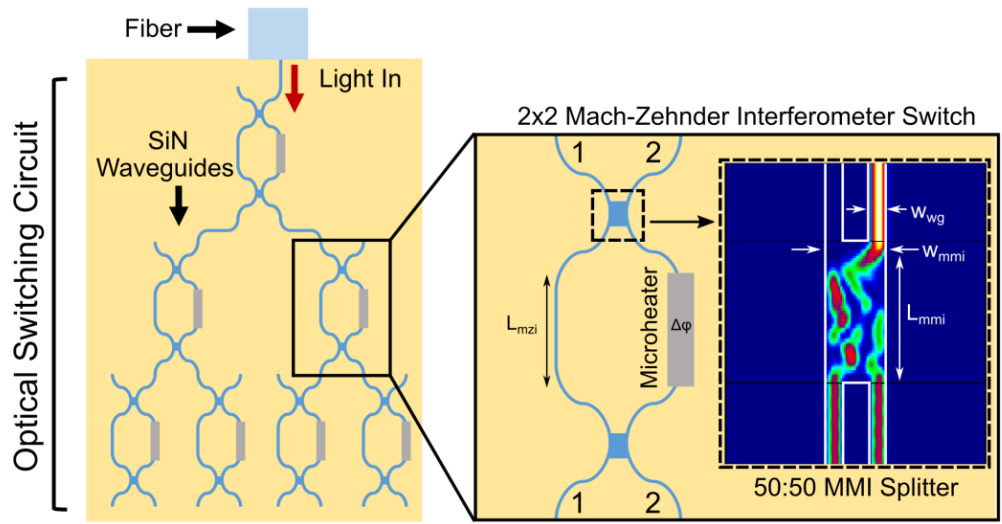

***Supp 1 Thermo-optic Switch Design.*** *The thermo-optic switch is based on a Mach-Zehnder Interferometer (MZI) created using SiN waveguides and a Pt microheater. The light comes into the MZI into either arm with the same phase, the light is split equally at a 50:50 splitter, the light travels through both arms for the same length, and the light is split equally again with a 50:50 splitter. One of the arms has a microheater aligned on top, so as the power on the heater is increased, the refractive index of the waveguide in that arm is increased due to the thermo-optic effect. The change in refractive index leads to a phase difference (Δφ) between the two arms, which allows the continuous control of light between the two arms. Where Δφ is proportional to the change in index and $L_{mzi}$. For the splitter we use a multimode interferometer (MMI), with a length ($L_{mmi}$) of 12.6 μm and width ($w_{mmi}$) of 1.2 μm. An MMI uses higher order modes and the self-imaging principle to create fabrication tolerant, compact, and fairly polarization insensitive structures for splitting light. The inset shows a numerical simulation (Eigenmode Expansion Method, FIMMWAVE software) of the normalized intensity of the light when light is input through port 2. The structure behaves identically, if light is input through the port 1 instead.*

- ■ Si₃N₄
- ■ SiO₂
- ■ Si
- ■ Pt

1) Deposit low pressure chemical vapor silicon nitride on oxidized silicon wafer.

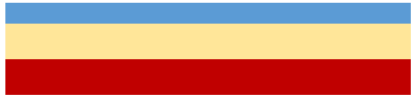

2) Pattern and etch silicon nitride waveguides.

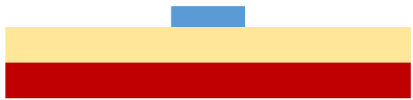

3) Deposit low pressure chemical vapor silicon dioxide.

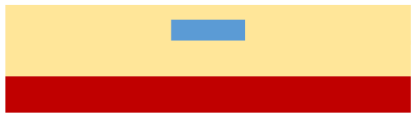

4) Sputter deposit and pattern platinum heaters using lift-off.

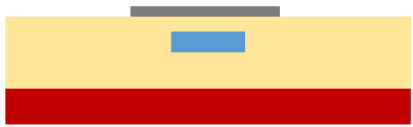

5) Dice chips to form probe.

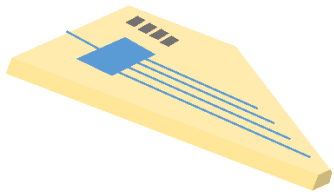

6) Cover the base of the backside of the chip with a protective polymer and etch the backside silicon substrate.

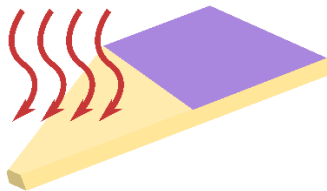

7) Mount chip on Printed Circuit Board (PCB), wire-bond chip to PCB, solder connector.

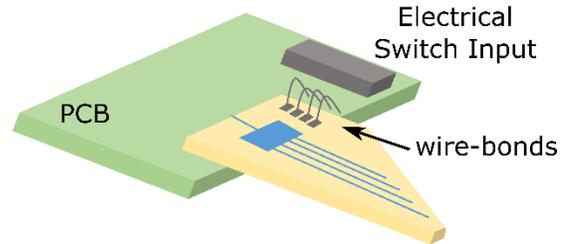

8) Cover wire-bonds with protective epoxy layer.

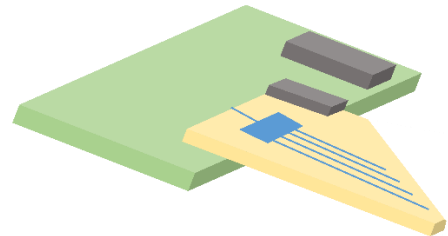

9) Actively align and attach pre-assembled fiber using UV-cure adhesive.

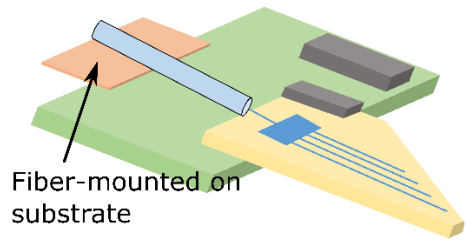

10) Attach pre-aligned electrodes on top of chip.

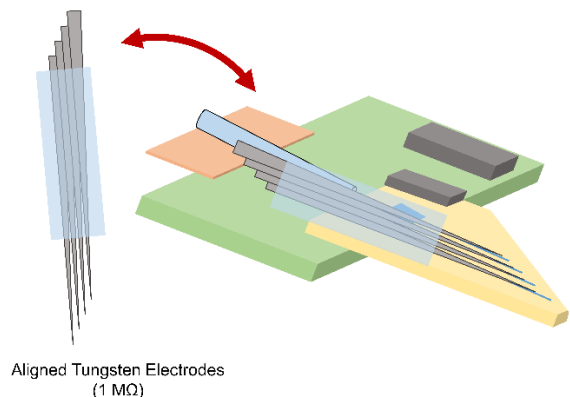

Aligned Tungsten Electrodes (1 MΩ)

*Supp 2 Fabrication and packaging process of the nanophotonic probe.*

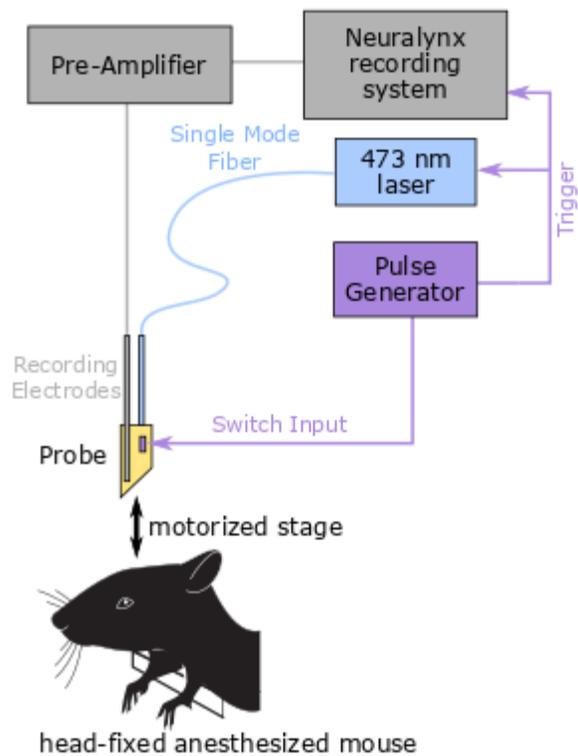

***Supp 3 Experimental setup for In-vivo demonstration.*** *The anesthetized mouse is placed onto the* stereotaxic *apparatus. A pulse generator (Pulse Pal) was used to send 7 timed voltage patterns to the 7 nanophotonic switches and 1 synchronized TTL output to trigger laser pulses. A Neuralynx Digital Lynx system was used to acquire the pre-amplified 8-channel neural activities. This system also included a digital input channel to acquire laser pulses sequence for synchronization of the light stimulus and neural spikes.*

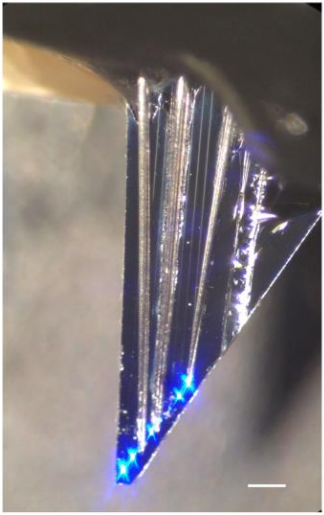 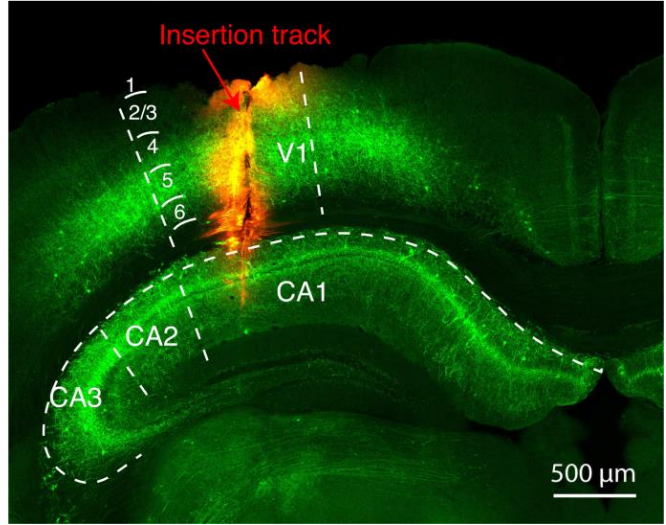

***Supp 4 Electrode arrangement and switching configuration for in-vivo experiment in Figure 4.*** *a) Shows a microscope image of the electrode arrangement of the device used in the in-vivo demonstration. Here we use three stereotrodes (pre-assembled pairs of tungsten electrodes) aligned to beam 1, beam 2, and beam 3 as shown in Figure 4a (scale bar is 250 µm). b) Histological section of the brain depicts the insertion track (coated with a dye (DiL) in red) across the layer 2-6 visual cortex (V1) and hippocampus (CA1), with the deepest site reaching ~1300 µm from the brain surface. ChETA-eYFP expressing in the Gad2 interneuron is shown in green.*

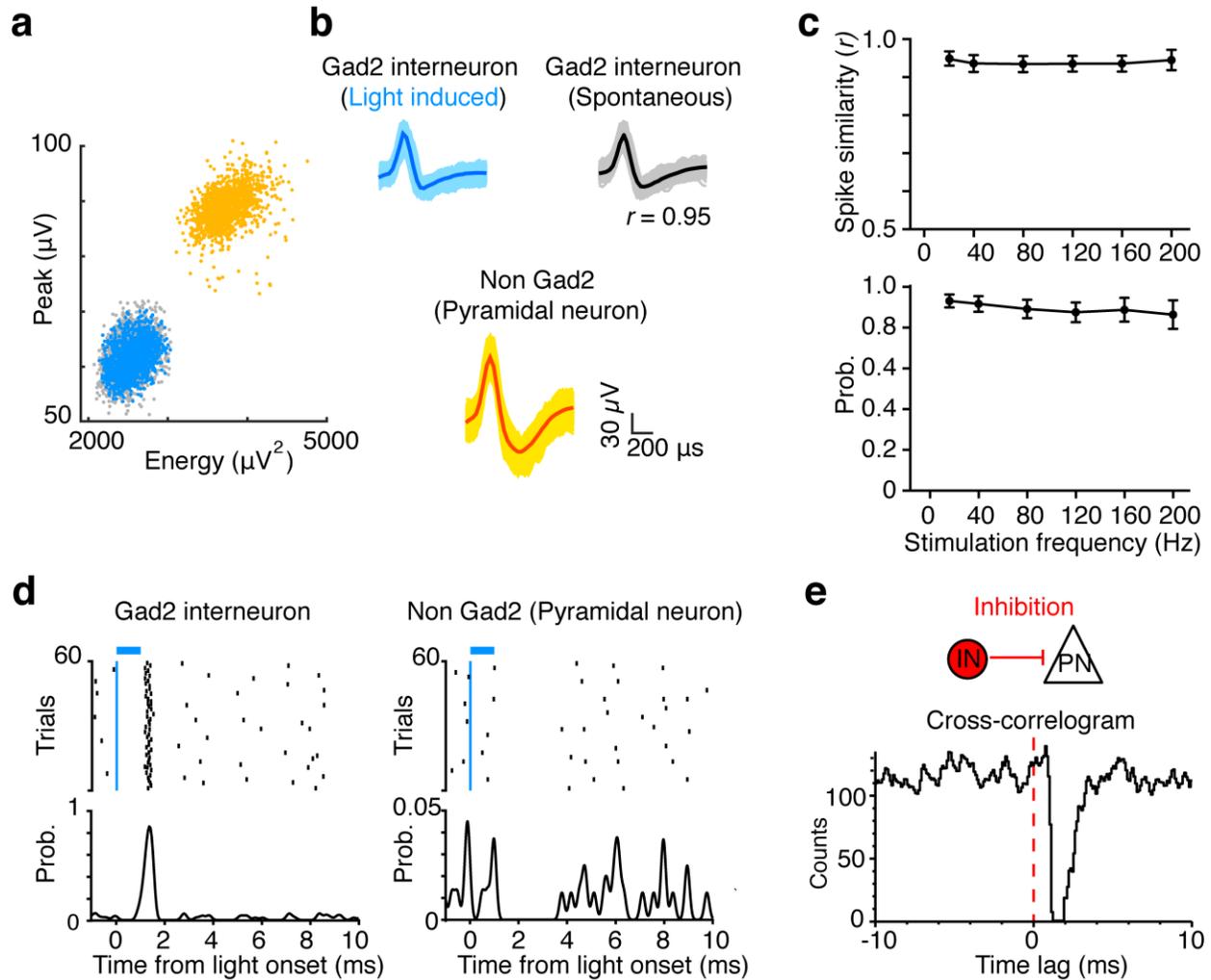

***Supp 5 Identification of ChETA-expressing Gad2 interneuron by light tagging test.*** *a) Spike sorting example. Two individual neuron clusters, recorded from the same recording electrode, are classified based on the property of spike waveform (energy and peak). ChETA-expressing Gad2 interneuron exhibits light-induced spikes (superimposed in blue dots), which are overlaid with the spontaneous spikes (grey dots). Non Gad2 putative pyramidal neuron doesn't express ChETA, thus exhibits only spontaneous spikes (yellow dots). b) The correspondent spike waveforms of Gad2 interneuron (light induced spikes in blue, and spontaneous spikes in grey; spike waveform similarity: r = 0.95) and putative non Gad2 pyramidal neuron (yellow). c) Summary of spike waveform similarity (r) of averaged light induced spikes and spontaneous spikes (top), and summary of light induced spike probability (bottom) throughout multiple stimulation frequencies (mean ± s.d., n=13, sorted neurons from 4 mice). d) Spike raster and peri-stimulus time histogram of Gad2 (left) and non Gad2 (left) neuron, with all spikes events aligned to light onset. Blue line indicates 1 ms light pulse. Only light tagged Gad2 interneuron exhibits low latency and high probability of light induced spikes. e) The correspondent cross-correlogram (CCG) between identified two neurons. Reference event: spikes from Gad2 interneuron (IN in red circle). Note strong and short-latency suppression of target spikes from non Gad2 pyramidal neuron (PN in white triangle), indicating the direct inhibitory interactions between neuron pairs.*

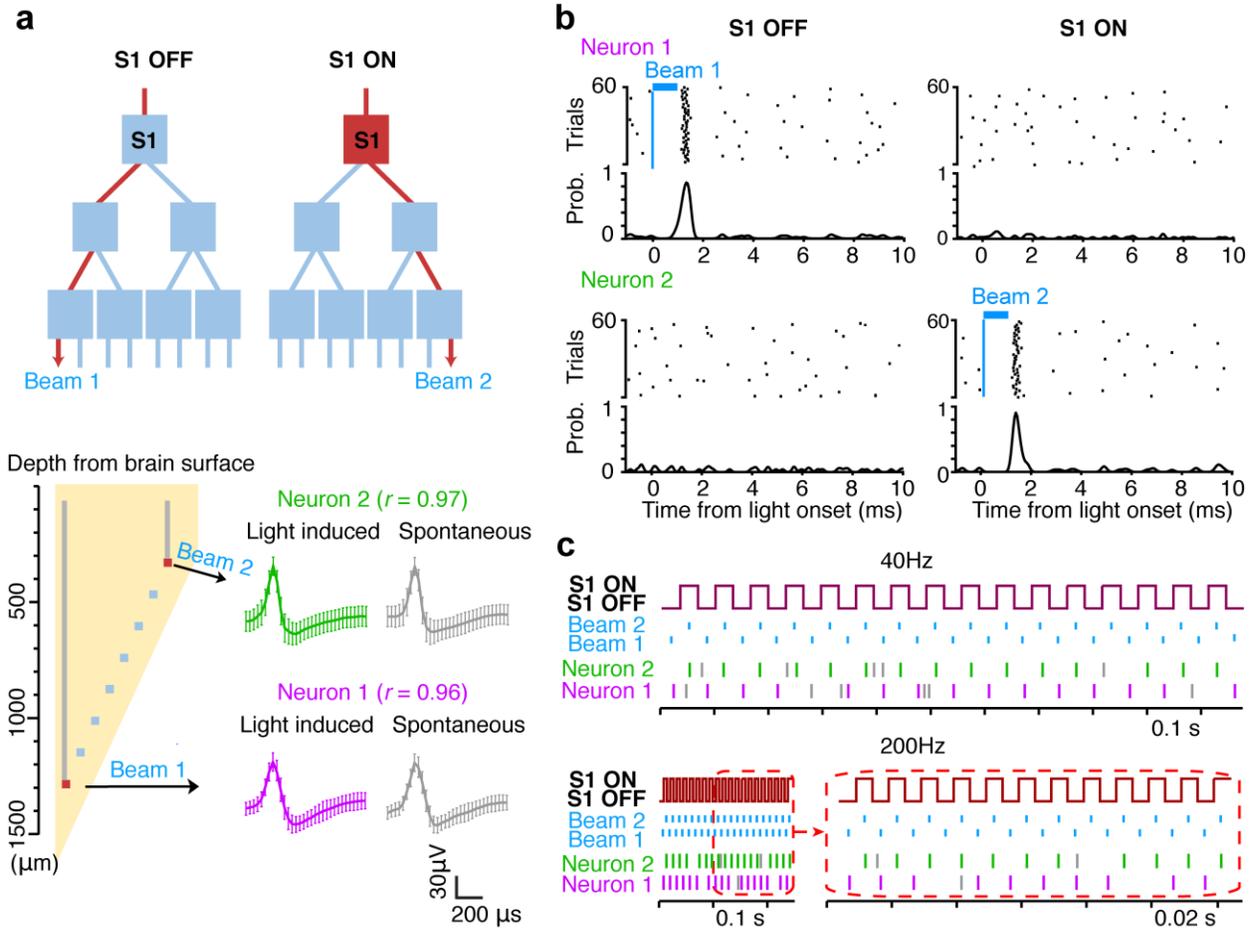

***Supp 6 In-vivo characterization of single switch performance (S1).*** *a) Top: switch diagram shows how S1 state (OFF or ON) determines the beam direction (beam 1 or beam 2). Bottom: schematic of beam and electrode arrangement for in-vivo examination of effective optogenetic control of ChETA-expressing Gad2 interneuron's activity across visual cortex (V1) and hippocampus in anesthetized mice. Two tungsten electrodes (grey lines) are positioned next to beam 1 and 2 (red squares separated in 875 μm in depth), respectively. From each recording site, single ChETA-expressing Gad2 interneuron (Neuron 1-2) is confirmed by evaluating the spike waveform similarity (r) of light induced spikes (averaged spike waveform in colored bold line, superimposed with s.d. of individual spikes in colored error bars) and spontaneous spikes (averaged spike waveform in grey bold line, superimposed with s.d. of individual spikes in grey error bars). b) Representative spike raster and peri-stimulus time histogram of Neuron 1 and 2 when S1 is ON and OFF (Light pulse width: 1ms). c) Spike raster of Neuron 1 and 2 which follow the S1 OFF (beam 1 on) or S1 ON (beam 2 on) state with strong fidelity at both 40 Hz and 200 Hz. For each neuron, the colored line indicates light induced spike, and the grey line indicates spontaneous spike.*

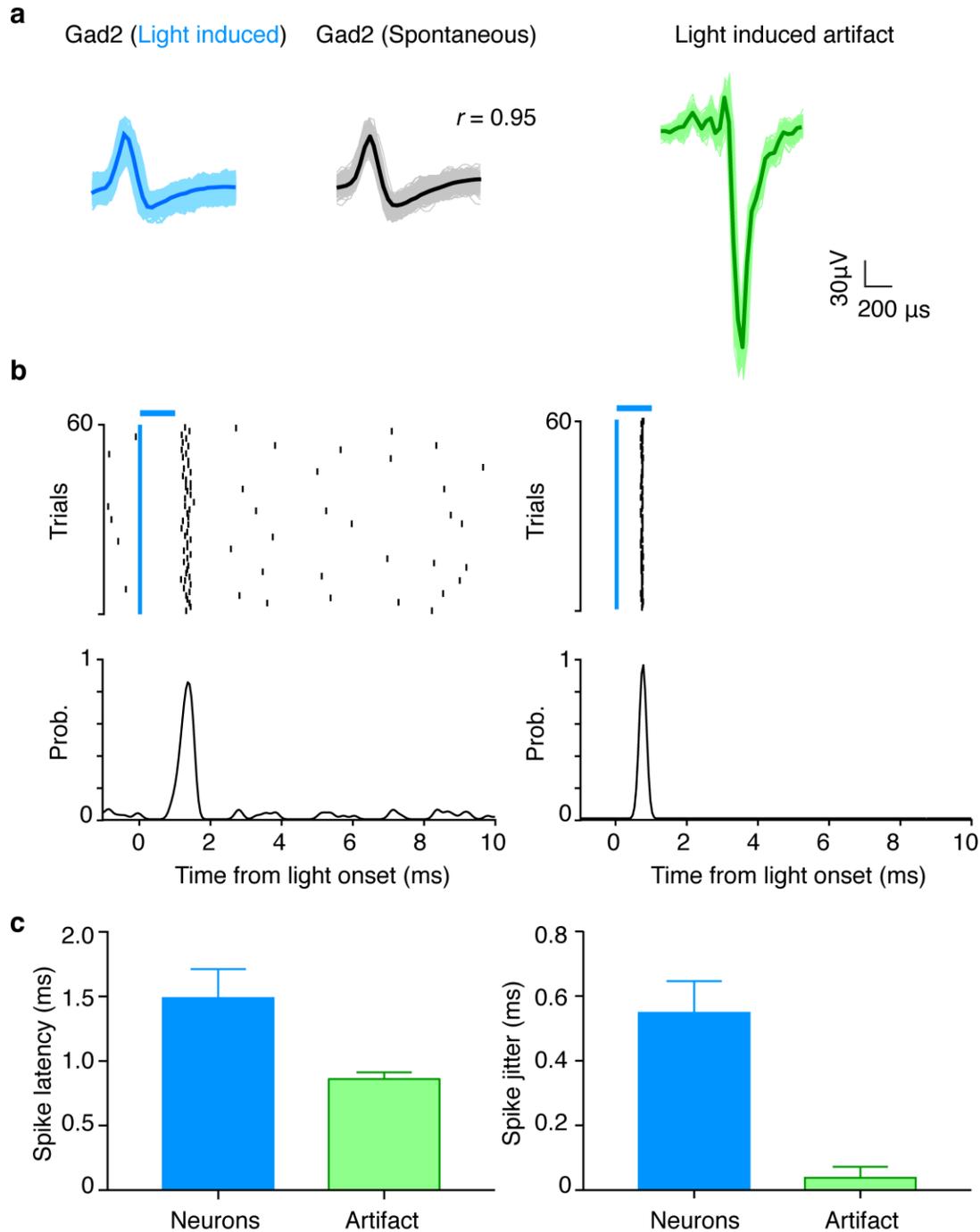

***Supp 7 Characterization of the difference in waveform and temporal response of light-induced spike and light-induced artifact.*** *We eliminated the light-induced artifact that commonly occurs when light is incident on the exposed tungsten electrode tip. Here we show, how we can distinguish this waveform through spike-sorting to recover neural spike information. a) Left, example of spike waveform taken from ChETA-expressing Gad2 interneuron in **Supp 5** (light induced spikes in blue, and spontaneous spikes in grey). The exposed tip of tungsten electrode is placed < 50 μm next to the emitter, presumably out of the cone of light; Right, example of robust and repeated light-induced artifact observed on tungsten electrode tip immersed in saline, which is directly exposure to the cone of light beam. No event matches with light-induced artifact's waveform when light is off. b) Raster plot and peri-stimulus time histogram of light-induced spikes and light-induced artifacts. Blue line indicates 1 ms light pulse and pulse onset is set to time '0'. c) Comparison of latency and jitter of the light induced spikes and artifacts. Optical artifact exhibits much shorter latency and much lower jitter than light-induced spike.*